\documentclass[prl, showpacs, twocolumn, floatfix]{revtex4}

\usepackage{graphicx}
\usepackage{amsmath, amsfonts, amssymb, bm}
\usepackage[]{psfrag}

\psfragscanoff
\usepackage{pifont}
\usepackage{dcolumn}
\usepackage{slashed}

\newcommand{\wg}{\omega_{\gamma}}

\begin{document}

\title{Strongly enhanced pair production in combined high- and low-frequency laser fields}

\author{Martin J. A. Jansen}
\author{Carsten M\"uller}
\affiliation{Institut f\"ur Theoretische Physik I, Heinrich-Heine-Universit\"at D\"usseldorf, Universit\"atsstr. 1, 40225 D\"usseldorf, Germany}

\date{\today}

\begin{abstract}
Production of electron-positron pairs by a high-energy probe photon propagating through a high-intensity laser field is considered in the nonperturbative interaction regime. The laser field consists of a strong low-frequency and a weak high-frequency component. While for each component alone photoproduction of pairs is strongly suppressed, we show that their combination can largely amplify the pair production probability by facilitating to bridge the energetic barrier of the process. Our predictions can be tested by utilizing presently available high-intensity laser devices.
\end{abstract}


\maketitle

Decay of the quantum vacuum into pairs of electrons and positrons in the presence of a very strong electromagnetic field belongs to the most intriguing predictions of quantum electrodynamics (QED). For the case of a static electric field, which was first examined in the 1930s \cite{Sauter}, the process is nowadays often referred to as the Schwinger effect \cite{Schwinger,Schwinger2}. The corresponding pair production rate has the form $\mathcal{R}\sim \exp(-\pi E_{\rm cr}/E)$, where $E$ is the applied field and $E_{\rm cr}=m^2/e$ the critical field of QED. Here, $e$ and $m$ are the positron charge and mass, respectively, and we use relativistic units with $\hbar = c = 1$. Intuitively, the exponential rate dependence may be understood by viewing the pair production as quantum mechanical tunneling through a potential barrier of height $\sim 2m$ and width $\ell\sim 2m/eE $. The barrier arises from the field-induced tilting of the positive- and negative-energy states in the Dirac sea picture. The Schwinger 
effect has a manifestly nonperturbative character due to its non-analytic field dependence. An experimental observation has so far been prevented by the huge value of $E_{\rm cr}\approx 10^{16}$\,V/cm which is not accessible in the laboratory.

Pair production processes resembling the Schwinger effect have also been predicted for space-time dependent fields, such as provided by high-intensity laser sources, for instance. While a single plane-wave laser field cannot create pairs due to constraints from energy-momentum conservation, pair production can be induced by an additional probe photon $\omega_\gamma$ traveling through a strong laser wave \cite{Reiss, Ritus, Fofanov, Serbo, Kampfer, Kasia, Selym, review2}. When the field frequency $\omega$ is relatively small, so that the dimensionless parameter $\xi\equiv eE/m\omega\gg 1$, the rate of this strong-field Breit-Wheeler process is \cite{Reiss, Ritus}
\begin{eqnarray}
\label{ExpRate}
\mathcal{R} &=& \frac{3^2\alpha m^2}{2^7\sqrt{2\pi}\omega_\gamma} \left(\frac{E'}{E_{\rm cr}} \right)^{3/2}\! \exp\!\left(-\frac{8}{3}\frac{E_{\rm cr}}{E'}\right)
\end{eqnarray}
where $\alpha=e^2$ denotes the finestructure constant and $E'=\frac{2\omega_\gamma}{m}E\ll E_{\rm cr}$ the laser field amplitude transformed to the center-of-mass frame of the created pair \cite{review2}. The similarity between \eqref{ExpRate} and the Schwinger rate is due to the fact that the characteristic length scale of pair production is much shorter than the laser wavelength, so that the field appears as quasi-static during the process. 

Conversely, at high laser frequencies with $\xi\ll 1$, pairs can be created dynamically by absorbing the probe photon together with a certain number $n$ of laser photons, according to $\omega_\gamma + n\omega \to e^+e^-$. In this regime, the rate scales perturbatively as $\mathcal{R}\sim \xi^{2n}$. In the 1990s, the strong-field Breit-Wheeler process was observed in a pioneering experiment at SLAC (Stanford, California) operating at $\xi\approx 0.4$ \cite{SLAC, Huayu}. However, pair production in the fully nonperturbative regime as described by the Schwinger-like rate \eqref{ExpRate} has not yet been observed.

Due to the fundamental importance of Schwinger(-like) pair production for our understanding of nonperturbative quantum field theories,  schemes to facilitate its observation have been developed in recent years. A dynamically assisted variant has been put forward where a rapidly oscillating electric field is superimposed onto a slowly varying electric field \cite{Schutzhold}. Pair production in this field combination was shown to be strongly enhanced, while preserving its nonperturbative character \cite{Schutzhold, Alkofer}. Similar enhancement effects were obtained for pair production in spatially localized fields \cite{Grobe} and in relativistic proton-laser collisions \cite{Antonino}. Experimental verification of these interesting predictions is hindered by the fact, though, that the fields employed are difficult to generate \cite{Schutzhold, Alkofer, Grobe} or a very powerful proton accelerator is required \cite{Antonino}. For the strong-field Breit-Wheeler process, which -- according to the successful 
SLAC experiment \cite{SLAC} -- appears most accessible to observation, corresponding enhancement schemes have not been examined yet \cite{Kampfer2}.

The significance of the Schwinger effect is further underlined by the fact that analogies of it are seeked in various areas of physics. They have been identified in a variety of systems such as graphene layers in external electric fields \cite{graphene} as well as ultracold atom dynamics \cite{cold} and light propagation \cite{waveguide} in optical lattices. In these systems, rather than electron-positron pair creation, the transition probability between quasiparticle and hole states exhibits Schwinger-like properties.

In this Letter, we study nonperturbative Breit-Wheeler pair production in a laser field consisting of a strong low-frequency and a weak high-frequency component. An $S$ matrix calculation within the framework of laser-dressed scalar QED employing Gordon-Volkov states is carried out \cite{scalar}. Our method allows us to probe a wide range of field frequencies, lying far from or close to the reaction threshold. Absorption of one or more photons from the high-frequency laser mode is considered. We show that, due to a favorable combination of the quasi-static and dynamical production mechanisms, the pair yields can be largely enhanced for experimentally available field parameters.

The bifrequent laser field is composed of two linearly polarized modes, with frequencies $\omega_j$, uniform propagation direction ${\bf n}$ and perpendicular polarization vectors ${\bf e}_j$ ($j=1,2$). It is described by the classical vector potential in radiation gauge
\begin{eqnarray}
\label{laserfield}
 {\bf A}_L={\bf A}_1+{\bf A}_2\  , \ {\bf A}_j(\eta_j)=a_j \cos(\eta_j-\varphi_j)\,{\bf e}_j
\end{eqnarray}
with the phases $\eta_j=(k_j x)=\omega_j (t - {\bf n}\cdot {\bf r})$, space-time coordinate $x^\mu=(t,{\bf r})$, wave four-vectors $k_j^\mu=\omega_j(1,{\bf n})$ and phase shifts $\varphi_j$. The potential amplitudes $a_j$ can also be measured by the Lorentz scalars $\xi_j=\frac{ea_j}{m}$.

The Gordon-Volkov states in the field \eqref{laserfield} for a particle (antiparticle) with free four-momentum $p_-^\mu=(E_{p_-},{\bf p}_-)$ [$p_+^\mu=(E_{p_+},{\bf p}_+)$] can be written as \cite{IZ, review1}
\begin{eqnarray}
 \Phi_{p_\pm}(x)=\frac{1}{\sqrt{2E_{p_\pm}V}}\, e^{i \left[ \pm (p_\pm x) - \Lambda_1 - \Lambda_2 \right]}
\end{eqnarray}
with $\Lambda_j = \frac{1}{(k_jp_\pm)} \int^{\eta_j} [e {\bf p}_\pm \cdot {\bf A}_j(\tilde{\eta}) \mp e^2 {\bf A}^2_j(\tilde{\eta}) ] d\tilde{\eta}$ and a normalization volume $V$.
They contain the interaction of the particles with the laser field to all orders and give rise to the laser-dressed momenta $q_\pm^\mu=p_\pm^\mu+2 k_1^\mu y_1^\pm + 2 k_2^\mu y_2^\pm $ and laser-dressed mass $m_*= m [{1+\frac{1}{2}(\xi_1^2+\xi_2^2)}]^{1/2}$. Here the abbreviation $y_j^\pm = \frac{e^2a_j^2}{8 (k_jp_\pm)}$ has been used.

The probe photon is treated as a mode of the quantized radiation field which has wave four-vector $k_\gamma^\mu = (\omega_\gamma,{\bf k}_\gamma)$, polarization vector ${\boldsymbol\epsilon}_\gamma$, and mode index $\lambda_\gamma$. 
Its absorption during the process is described by the field operator
\begin{eqnarray}
 {\hat {\bf A}}_\gamma=\sqrt{\frac{2 \pi}{V \wg}}\, e^{-i(k_\gamma x)}\,{\boldsymbol \epsilon}_\gamma\, {\hat c}_{{\bf k_\gamma}\lambda_\gamma}\ ,
\end{eqnarray}
accordingly, with the annihilation operator ${\hat c}_{{\bf k_\gamma}\lambda_\gamma}$.

Within laser-dressed scalar QED, the $S$ matrix for pair production in the considered field configuration reads \cite{review1}
\begin{eqnarray}
\label{Smatrix}
 S= -i \int_{-\infty}^{+\infty} dt\, \langle \Psi_{p_-} | {\hat H}_{\rm{int}} | \Psi_{p_+,\gamma}\rangle\ ,
\end{eqnarray}
with the initial state $|\Psi_{p_+,\gamma}\rangle=\Phi_{p_+}|{\bf k}_\gamma \lambda_\gamma\rangle$, final state $|\Psi_{p_-}\rangle=\Phi_{p_-}|0\rangle$, and the interaction Hamiltonian
\begin{eqnarray}
{\hat H}_{\rm{int}}=-ie \left( {\hat {\bf A}}_\gamma \cdot \overset{\rightarrow}{\boldsymbol{\nabla}} - \overset{\leftarrow}{\boldsymbol{\nabla}} \cdot {\hat {\bf A}}_\gamma \right) + 2e^2 {\bf A}_L \cdot {\hat {\bf A}}_\gamma\ .
\end{eqnarray}

The space-time integral in \eqref{Smatrix} can be performed analytically. To this end, the periodic terms in the integrand are expanded into Fourier series by virtue of the formula $e^{i(u\sin(\phi)+v\sin(2\phi))}=\sum_{n=-\infty}^\infty e^{in\phi} \tilde{J}_n(u,v)$  \cite{ReissBessel}. This can be done for each laser mode separately, as no cross terms between them arise due to the orthogonal polarization of the modes. Generalized Bessel functions occur here which are composed of ordinary Bessel functions via $\tilde{J}_n(u,v)=\sum_{\ell=-\infty}^{\infty} J_{n-2\ell}(u)J_\ell(v)$. 

This way, assuming the probe photon to counterpropagate the laser beams, we obtain
\begin{eqnarray}
 S &=& S_0 \sum_{n_1,n_2=-\infty}^\infty M_{n_1,n_2}\, e^{i(n_1\varphi_1+n_2\varphi_2)} e^{i(h_1+h_2)}  \nonumber\\
 & & \times (2\pi)^4 \delta^4 \left(q_- + q_+ - k_\gamma - n_1 k_1 - n_2 k_2 \right)
\end{eqnarray}
with $S_0=-ie m\sqrt{\pi/(2V^3E_{p_+} E_{p_-}\omega_\gamma)}$, $h_j=h_j^++h_j^-$, $h_j^\pm=x_j^\pm\sin(\varphi_j)+y_j^\pm\sin(2\varphi_j)$, and $x_j^\pm=\mp \frac{ea_j{\bf p}_\pm \cdot {\bf e}_j }{ (k_jp_\pm)}$ for $j=1,2$. The reduced matrix element is
\begin{eqnarray}
  M_{n_1,n_2} &=& \frac{({\bf p}_--{\bf p}_+)\cdot {\boldsymbol\epsilon}_\gamma}{m} \tilde{J}_{n_1}\tilde{J}_{n_2} \nonumber\\ 
& & +\, \xi_1 ({\bf e}_1 \cdot {\boldsymbol\epsilon}_\gamma) \left( \tilde{J}_{n_1+1} + \tilde{J}_{n_1-1} \right) \tilde{J}_{n_2} \nonumber \\
& & +\, \xi_2 ({\bf e}_2 \cdot {\boldsymbol\epsilon}_\gamma) \left( \tilde{J}_{n_2+1} + \tilde{J}_{n_2-1} \right) \tilde{J}_{n_1}
\end{eqnarray}
where the arguments of the generalized Bessel functions $\tilde{J}_{n_j}$ and $\tilde{J}_{n_j\pm1}$ read $u_j = -(x_j^++x_j^-)$, $v_j = -(y_j^++y_j^-)$.

Since we are interested in a situation where the laser frequencies are largely different, $\omega_1\gg\omega_2$, we shall assume that the values of $\omega_1$ and $\omega_2$ are not commensurate. In this case, the values of the phase shifts are immaterial and will be set to zero. The square of the $S$ matrix can be written as \cite{commensurate}
\begin{eqnarray}
\label{Ssq}
|S|^2 &=& (2\pi)^4 |S_0|^2\, VT \sum_{n_1,n_2} |M_{n_1,n_2}|^2  \nonumber\\
& & \times\, \delta^4 \left(q_- + q_+ - k_\gamma - n_1 k_1 - n_2 k_2 \right)
\end{eqnarray}
where the interaction space-time volume $VT$ stems from the square of the $\delta$ function. From \eqref{Ssq} the production rate  follows by dividing out the interaction time, integrating over the produced particle momenta, and averaging over the probe photon polarization states:
\begin{eqnarray}
 \mathcal{R} = \frac{1}{2}\sum_{\lambda_\gamma}\int \frac{V d^3p_+}{(2\pi)^3} \int\frac{V d^3p_-}{(2\pi)^3} \frac{|S|^2}{T}\ .
\end{eqnarray}
Since the energy-momentum balance in the $\delta$ function is given in terms of the laser-dressed momenta, it is convenient to rewrite the integrals by using the relation $\frac{d^3p_\pm}{E_{p_\pm}}=\frac{d^3q_\pm}{E_{q_\pm}}$ \cite{LL}. Moreover, for our purposes it shall be useful to decompose the total production rate into partial contributions $\mathcal{R}(n_1)$ associated with a certain number $n_1$ of high-frequency laser photons absorbed. We therefore express our final result in the form
\begin{eqnarray}
\label{R}
\mathcal{R} = \sum_{n_1} \mathcal{R}(n_1)\ ,
\end{eqnarray}
with
\begin{eqnarray}
\label{Rn1}
\mathcal{R}(n_1) &=& \frac{\alpha m^2}{16\pi\omega_\gamma}\sum_{n_2}\sum_{\lambda_\gamma}\int \frac{d^3q_+}{E_{q_+}} \int\frac{d^3q_-}{E_{q_-}} |M_{n_1,n_2}|^2 \nonumber\\
& & \times\,\delta^4 \left(q_- + q_+ - k_\gamma - n_1 k_1 - n_2 k_2 \right)\ .
\end{eqnarray}
The numbers of laser photons involved have to be large enough to satisfy the condition $(n_1 \omega_1 + n_2 \omega_2)\omega_\gamma \geq m_*^2$.

\begin{figure}[b]
\begin{center}
\resizebox{8cm}{!}{\includegraphics{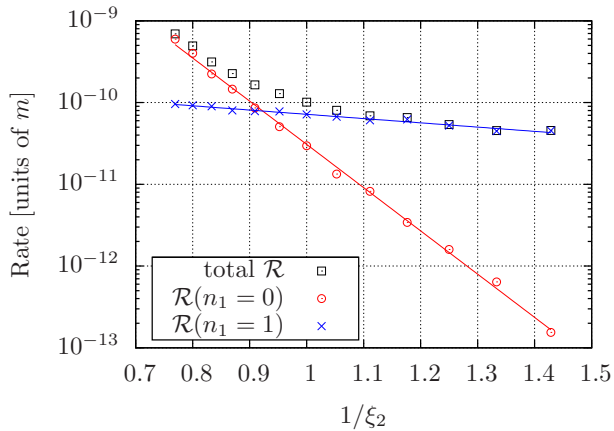}}
\caption{\label{n1one} (Color online) Rate for pair production by a high-energy probe photon with $\omega_\gamma\approx 0.9487m$ counterpropagating a bifrequent laser beam with $\omega_1=\omega_\gamma$ (such that $\Delta = 0.1$), $\xi_1=2\times 10^{-3}$ and $\omega_2 = 0.1m$, as a function of the inverse field intensity parameter $\xi_2$ (black squares). The red circles and blue crosses show the respective contributions to the rate from absorption of zero and one assisting $\omega_1$-photon, as indicated. The lines are exponential fit functions to the data (see text for further details). Note that $m=7.76\times 10^{20}$\,s$^{-1}$ in SI units.}
\end{center} 
\end{figure}

Based on Eqs.~\eqref{R} and  \eqref{Rn1} we have calculated the rate for strong-field Breit-Wheeler production of (scalar) electron-positron pairs in a laser field composed of a weak high-frequency mode ($\xi_1\ll 1$, $\omega_1\sim m$) and a strong low-frequency mode ($\xi_2\sim 1$, $\omega_2\ll m$). Figure~\ref{n1one} shows our results on the total rate $\mathcal{R}$ and the relevant partial rates $\mathcal{R}(n_1)$. 

Let us first consider the partial rate which does not involve photon absorption from the first mode (red circles). Note that, for the chosen parameters, it practically coincides with the rate for pair production by the probe photon in the presence of the strong low-frequency mode alone. To a good approximation the numerical data exhibit an exponential dependence on the laser field strength which can be fitted to the curve $f(\xi_2)\sim\exp(-\Gamma/\xi_2)$, with $\Gamma \approx 12.2$. Despite the moderate values of $\xi_2$, the dependence is in reasonable agreement with Eq.~\eqref{ExpRate} which gives $\Gamma\approx 14.1$. That Schwinger-like rate dependences are obtained already for intensity parameters starting from $\xi\approx 1$ is known from other laser-induced pair production processes as well \cite{xi1}. 

Participation of the weak high-frequency mode can enhance the pair production rate tremendously, in spite of its low intensity. Indeed, for the most part displayed in Fig.~\ref{n1one}, the total rate is dominated by the channel which involves absorption of one high-frequency laser photon \cite{n2zero}. The corresponding partial rate (blue crosses) still exhibits a nonperturbative exponential dependence on the strong field, however, with a modified and largely reduced exponent $\tilde\Gamma\approx 1.20$. 
We note that the typical number $n_2$ of absorbed low-frequency laser photons ranges from about 5 to 30 here, whereas for the non-assisted process from approximately 15 to 25.

The physical origin of the enhancement effect and the functional form of the rate can be understood as follows. A collision of two photons $k_1$ and $k_\gamma$ would allow for pair production by the usual Breit-Wheeler process if the center-of-mass energy fulfills $(k_1+k_\gamma)^2\ge 4m^2$. In the present case, however, the c.m. energy lies below the threshold. It is natural to introduce, accordingly, a gap parameter $\Delta$ by $(k_1+k_\gamma)^2 = 4m^2(1-\Delta)$. For counterpropagating photon beams, one obtains
\begin{eqnarray}
\label{Delta}
\Delta = 1-\frac{\omega_1\omega_\gamma}{m^2}\ .
\end{eqnarray}
In the c.m. frame, this may be rewritten as $\Delta = \delta (m+\omega_1)/(2m^2)$, where $\delta = 2(m-\omega_1)$ represents the remaining energy gap for pair production which has to be bridged by the influence of the strong low-frequency laser component. In order to gain an intuitive understanding we replace the latter by a constant electric field ${\bf E} = E_0{\bf e}_2$ which may induce a tunneling through the $\delta$ barrier. The corresponding tunneling rate is of the form $\exp(-\mathcal{G})$ with the Gamow factor (see also \cite{Antonino})
\begin{eqnarray}
\label{Gamow}
\mathcal{G} = 2\int_0^\ell\!\! \sqrt{2m(\delta - eE_0y)}\,dy 
= \frac{4\sqrt{2}}{3}\frac{E_{\rm cr}}{E_0}\left( \frac{2m\Delta}{m+\omega_1}\right)^{\!3/2}
\end{eqnarray}
where the tunneling length $\ell$ follows from $eE_0\ell = \delta$. Equation\,\eqref{Gamow} conforms with the exponential found for the blue line in Fig.~\ref{n1one} rather well [$\mathcal{G}\approx 0.62/\xi_2$].

\begin{figure}[t]
\begin{center}
\resizebox{8cm}{!}{\includegraphics{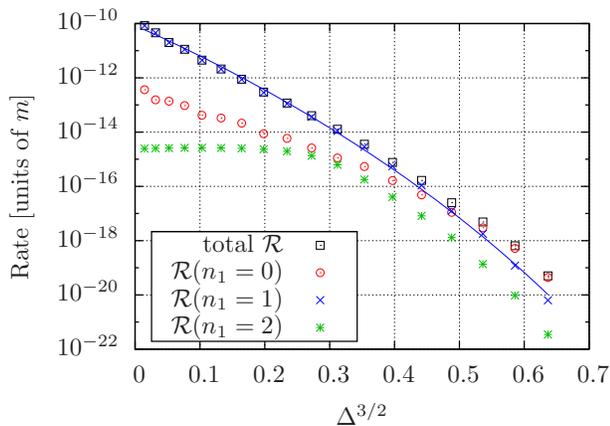}}
\caption{\label{DeltaFig} (Color online) Dependence of the pair production rate on the gap parameter at $\omega_1=\omega_\gamma=m\sqrt{1-\Delta}$, $\xi_1=2\times 10^{-3}$, $\omega_2 = 0.1m$, and $\xi_2=0.7$ (black squares). The red circles, blue crosses and green asterisks show the respective contributions to the rate from absorption of zero, one and two assisting $\omega_1$-photons, as indicated. The blue line is a fit function of the form suggested by Eq.~\eqref{Gamow}.}
\end{center} 
\end{figure}

The enhancement effect due to the weak high-frequency mode is most pronounced when the low-frequency component is not too strong. When the latter exceeds a certain level, the effect of the dynamical enhancement diminishes and the dominant mechanism of pair production goes over into the usual strong-field Breit-Wheeler process [see Eq.~\eqref{ExpRate}]. The $\xi_2$ value where this transition occurs can be estimated from the relation $\xi_1^2\exp(-\tilde\Gamma/\xi_2)\approx \exp(-\Gamma/\xi_2)$. It leads to $\xi_2\approx 1$, in accordance with Fig.~\ref{n1one}. The perturbative factor of $\xi_1^2$ in the above relation arises from the participation of one high-frequency photon $\omega_1$ in the dynamically assisted process, since $\mathcal{R}(n_1)\sim\xi_1^{2n_1}$ for $\xi_1\ll 1$.

Figure~\ref{DeltaFig} shows the dependence of the production rate on the gap parameter over a wide range. When $\omega_1\gtrsim 0.65m$, the pair production is dominated by the channel which involves absorption of one high-frequency laser photon. The corresponding partial rate can be well described by a fit function based on Eq.~\eqref{Gamow}. In particular, for small values of $\Delta$, the rate behaves as $\mathcal{R}\sim\exp(-\kappa\Delta^{3/2})$ with $\kappa\approx 29.5$ whereas Eq.~\eqref{Gamow} predicts $\kappa\approx 26.9$. Note that in this region the partial rate involving no high-frequency laser photon exhibits a step-like form which results from multiphoton channel closings. For comparison, Fig.~\ref{DeltaFig} also shows the partial rate involving two $\omega_1$-photons which, for the chosen parameter set, is surpressed. It saturates at $\Delta\lesssim 0.38$ where the corresponding threshold is exceeded, i.e., $2\omega_1\omega_\gamma>m_*^2$.

\begin{figure}[t]
\begin{center}
\resizebox{8cm}{!}{\includegraphics{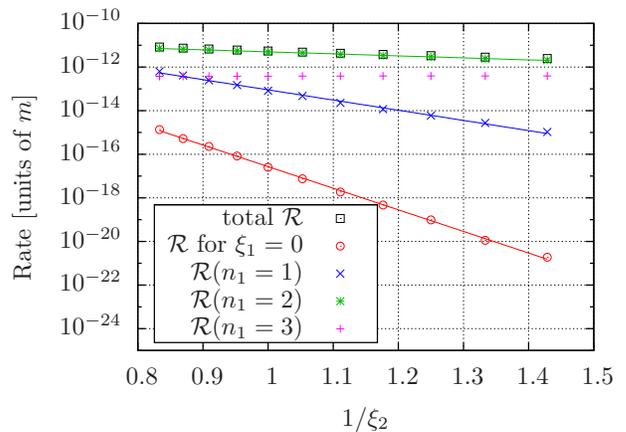}}
\caption{\label{n1two} (Color online) Rate for pair production by a high-energy probe photon with $\omega_\gamma\approx 0.9487m$ counterpropagating a bifrequent laser beam with $\omega_1=\omega_\gamma/2$, $\xi_1=4.5\times 10^{-2}$ and $\omega_2 = 0.05m$, as a function of the inverse field intensity parameter $\xi_2$ (black squares). The blue crosses, green asterisks and pink daggers show the contributions to the rate from absorption of one, two and three assisting $\omega_1$-photons, respectively. The red circles show the production rate in the absence of the high-frequency laser mode. The lines are exponential fit functions to the data.}
\end{center} 
\end{figure}

In certain parameter regimes, the pair production can be dominated by channels involving the absorption of more than one photon $\omega_1$. We note in this context that the contributions from higher orders in the weak laser mode are naturally included within our Volkov-state approach [see Eq.~\eqref{R}] which complements previous treatments of dynamically enhanced pair production based on other methods \cite{Schutzhold,Alkofer,Grobe,Antonino}. An example is shown in Fig.~\ref{n1two}, where the dominant channel involves the absorption of two high-frequency laser photons. It leads to a rate enhancement of up to ten orders of magnitude in the considered range of field intensities. As before, the rate shows an exponential behavior, with $\tilde\Gamma \approx 2.12$. It closely corresponds to $\mathcal{G}\approx 1.88/\xi_2$ from Eq.~\eqref{Gamow}, using the appropriately modified gap parameter $\Delta = 1-2\omega_1\omega_\gamma/m^2$.

Our numerical results shown above refer to the reference frame where $\omega_\gamma$ equals $\omega_1$ or $2\omega_1$, respectively. The parameters of Fig.~\ref{n1one}, for example, could be realized by lab frame values of $\omega_2 \approx 2.4$\,eV (Nd-YAG laser) with $I_2 \approx 10^{19}$\,W/cm$^2$, $\omega_1 \approx 24$\,eV with $I_1 \approx 4\times 10^{15}$\,W/cm$^2$ \cite{XUV}, and $\omega_\gamma \approx 10$\,GeV. The $\gamma$-photon can be generated by synchrotron radiation or, as in \cite{SLAC}, by Compton backscattering off a highly relativistic electron beam. In this regard it is worth mentioning that laser wakefield accelerators are currently approaching electron beam energies of this order \cite{laser-acc,BELLA}, this way offering prospects for a compact all-laser setup to implement our scheme. Assuming a laser-accelerated beam of $10^9$ electrons with 12 GeV energy and a Compton conversion efficiency into $\gamma$-photons of $\sim\alpha\xi_1^2\omega_1\tau\approx 10^{-4}$, on the order of 0.1 
electron-positron pairs are created on average in the bifrequent laser field during an interaction time of $\tau\sim 100$\,fs.

In conclusion, dynamically enhanced pair production in a nonperturbative (Schwinger-like) regime was studied within a field configuration purely made of photons. We determined parameter domains where a weak high-frequency component superimposed onto an intense optical laser field leads to pronounced rate enhancement and identified transition regions between the competing pair production mechanisms. Our predictions can be tested by utilizing presently available laser technology.

Useful input by S. Villalba-Ch\'avez, R. Sch\"utzhold, and T. Toncian is gratefully acknowledged.
This study has been performed within the project MU 3149/1-1 funded by the German Research Foundation (DFG).


\begin{thebibliography}{100}
\bibitem{Kampfer2} The rate enhancement reported in \cite{Kampfer} has a different nature as it results from the broad frequency spectrum of a short laser pulse and applies to the perturbative regime.
\bibitem{scalar} The spin degree of freedom of the electron shall not be accounted for in our calculation. This simplification, which renders the physical content of the equations more transparent, is enabled by the fact that the production rates for scalar and fermion pairs coincide, apart from an overall numerical prefactor of order 0.1; see Eq.~(35) in \cite{Selym}.
\bibitem{n2zero} For comparison we note that a production rate of $2.5\times 10^{-15}m$ ($3.9\times 10^{-13}m$) would result in Fig.~\ref{n1one} (Fig.~\ref{n1two}), when the strong low-frequency laser mode was switched off.
\end{thebibliography}
\end{document}